\documentstyle[aps,preprint,epsfig,amssymb]{revtex}
\begin{document}
\draft


\title{
Production of $\bbox{\phi}$ Mesons in Near-Threshold
$\bbox{\pi\,N}$ and $\bbox{N\,N}$ Reactions
}


\author{
A.I.~Titov\,$^{a,b}$%
\footnote{E-mail address : {\tt atitov@thsun1.jinr.ru}},
B.~K\"ampfer$^a$%
\footnote{E-mail address : {\tt kaempfer@fz-rossendorf.de}}
and B.L. Reznik$^c$%
\footnote{E-mail address : {\tt reznik@dvgu.ru}},
}

\address{
$^a$Forschungszentrum Rossendorf, PF 510119, 01314 Dresden,
Germany\\
$^b$Bogoliubov Laboratory of Theoretical Physics, JINR,
Dubna 141980, Russia\\
$^c$Far-Eastern State University, Sukhanova  9, Vladivostok 690090,
Russia
}

\maketitle

\begin{abstract}
We analyze the production of $\phi$ mesons
in $\pi N$ and $NN$ reactions in the near-threshold region,
using throughout the conventional ``non-strange'' dynamics  based on such
processes  which are allowed by the non-ideal $\omega-\phi$ mixing.
We show that the occurrence of the direct $\phi NN$ interaction
may show up in different unpolarized and polarization
observables in $\pi N\to N\phi$ reactions.
We find a strong  non-trivial difference between
observables in  the reactions $pp\to pp\phi$ and $pn\to pn\phi$ caused
by the different role  of the spin singlet and triplet states
in the entrance channel. 
A series of predictions
for the experimental study of this effect is presented.\\[3mm]
key words: hadron reactions, phi production, threshold
\end{abstract}

\section{Introduction}

The present interest in the $\phi$ meson production in different
elementary reactions is related  to the strangeness degrees
of freedom in the nucleon.
Since the $\phi$ meson is thought to consist mainly of strange
quarks, i.e. $s\bar s$, with a rather small contribution of the light
$u$ and $d$ quarks, its   production
should be suppressed if the entrance channel does not possess
a considerable admixture of strangeness.
Indeed, the recent experiments on the proton annihilation at rest
(cf.\ \cite{Ellis} for references and a compilation of data) point to
a large apparent violation of the OZI rule~\cite{OZI},
which is interpreted \cite{Ellis} as a
hint to an intrinsic $s\bar s$ component in the proton.
However, the data can be
explained as well by modified meson exchange models \cite{LZL93}
without introducing any strangeness component in the nucleon or
OZI violation mechanisms. On the other hand,
the analysis of the $\pi N$ sigma term \cite{sigmaterm} suggests that
the proton might contain a strange quark admixture as large as 20\%.
Thus this issue remains  controversial.
Therefore it is tempting to look for other
observables \cite{Ellis,newexp,TOY97_98} that are sensitive to the
strangeness content of the nucleon. Most of them are related to
a possible  strong  interference of delicate $s \bar s$
knock-out and shake-off
amplitudes and the ``non-strange'' amplitude which is caused by OZI rule
allowed processes, or by  processes wherein the standard OZI rule
violation comes from the $\phi - \omega$ mixing.

As shown in Ref.~\cite{TOY97_98}, through this interference
the polarization observables
of the $\phi$ {\em photoproduction} process are
sensitive even to a rather small
strangeness admixture in the proton. However, the only
$^{3,1}S$ $s\bar s$ configurations may be seen in a such process.
The other configurations, such as ${}^{3,1}P$,
are suppressed by the selection rules and/or form factors.
Contrary to this,
Ellis et al. \cite{Ellis99} argue that the possibly
dominant $^{3}P_1$ configuration might be seen in {\em hadronic} reactions.

Obviously, reliable
information about the hidden strangeness
manifestation in the $\phi$ production
in $\pi p$ and $NN$ reactions can be obtained only when
the conventional, i.e. non-exotic, amplitude has been
understood quantitatively.
This is the objective of this work.
The dominant conventional processes  in $\pi N$ and $NN$ reactions
are depicted in Figs.~1 and 2, where
(a) is the mesonic exchange
process being allowed by the  finite
$\phi\rho\pi$ coupling strength and (b) is the direct $\phi$ radiation
from the nucleon legs,
which is proportional to the finite $\phi NN$ interaction strength.
It should be emphasized that the process 1a is a subprocess in the
diagram 2a, while the process 1b is a subprocess in the diagram 2b,
when the exchanged boson is a pion.

While the diagrams in Figs.~1 and 2 look like usual Feynman diagrams it
should be stressed they give a guidance of how to obtain from an
interaction Lagrangian of hadronic  fields a covariant parameterization
of observable in strict tree level approximation. Additional ingredients
are needed to achieve an accurate description of data within such a framework.
In particular, the vertices needs to be dressed by form factors.

The early theoretical studies~\cite{KO,TKS}
show, indeed, that predictions for hadronic observables are very sensitive
to the parameters of the monopole form factors
which can not be fixed unambiguously without adjustments
relying on the corresponding experimental data.
In our case one can rely on the recent measurement of the
ratio of the total cross
sections of $\phi$ and $\omega$ production in $pp$ reactions
studied by the DISTO Collaboration at
SATURNE at ${\rm T_{lab}}=2.85$ GeV
as well as on the $\phi$ angular
distribution in the $pp\to pp\phi$ reaction \cite{DISTO}
and on the total cross section \cite{DISTO99}.

In the (sub)threshold region also in heavy-ion reactions the $\phi$ production
data is accessible via the $K^+ K^-$ decay channel studied
with the $4\pi$ detector FOPI at SIS in GSI/Darmstadt \cite{Kotte}.
However, here an understanding of elementary hadronic reactions serves as
a prerequisite for interpreting the data. Rather the upcoming
proton and pion beam experiments with the HADES detector system at
SIS in GSI \cite{HADES} offer a chance to enlarge the data base. 
In particular,
HADES can identify the $\phi$ also via the $e^+ e^-$ channel.

Finally we mention that for pion-induced reactions at the proton also
near-threshold data for the $\phi$ production are available \cite{DATApi}.

An important step towards an understanding of the structure
of the ``non-strange'' $pp\to pp\phi$ reaction mechanisms
was made recently \cite{Nakayama99}.
The focus of Ref.~\cite{Nakayama99}
is the determination of the parameters of
the direct $\phi NN$ interaction, 
thus reducing the above mentioned uncertainties,
by analyzing the combined
the $pp\to pp \omega$  and $pp\to pp \phi$
reactions and the corresponding DISTO data \cite{DISTO}
at a given beam energy;
for the reaction $pp\to pp\omega$ just the same mechanism
is assumed as those for $pp\to pp\phi$, as shown in Fig.~2.
Clearly, at one given beam energies the excess energy for both
reactions are quite different.

In this paper we therefore attempt a different approach with a similar goal.
We reduce the uncertainties
of the reaction mechanism by a combined study of the to each other related
reactions $pp\to pp\phi$ and  $\pi^-p\to n\phi$
using the known data within the same interval of excess energies
of $20-100$ MeV~\cite{DISTO99,DATApi}.
For the absolute normalization
of the angular distribution in $pp\to pp\phi$ \cite{DISTO} we use the
recently published
total cross sections of the reaction $pp\to pp\phi$ \cite{DISTO99}.
It turns out that this value is compatible, within given error bars,
with an extrapolation of the previously measured ratio of
the total cross sections of $pp\to pp\phi$ to $pp\to pp\phi$
reactions, $\sigma_\phi/\sigma_\omega=(3.7\pm0.5)\times10^{-3}$ at
$T_{lab}=2.85$ GeV \cite{DISTO}, by normalizing it
to the old bubble chamber data for $\sigma_\omega$ at various excess
energies \cite{Hibou}.
The extrapolated value of the cross section differs from the new
value \cite{DISTO99} thus influencing to some extent
adjusted parameters.

In comparison with previous works we are doing the next step towards
an understanding of the dynamics of $\phi$ production in hadronic reactions.
We present a systematical analysis of
$\pi^-p\to n\phi$,  $pp\to pp\phi$  and $pn\to pn\phi$ reactions
in the near-threshold region where the destructive interferences
between the two mechanisms (a) and (b) in Figs.~1 and 2 are essential.
We are going to show that basically there are two different sets 
of the model parameters.
One of them corresponds to the case when the mesonic exchange
channel (a) is dominant for the $\pi^-p \to n\phi$ reaction (Fig.~1),
and in the second case the direct emission mechanism (b)
is dominant. For both sets of solutions we calculate the total
and differential cross sections and spin density matrix, responsible for the
$\phi\to e^+e^-$ and $\phi\to K^+K^-$ decay angular distributions
and show in which
observables the direct $\phi NN$ interaction might be clearly manifest.
We present also a combined  analysis of $pp\to pp\phi$ and $pn\to pn\phi$
reactions at a finite energy excess with taking into account
the final state interaction and analyze the deviation of predicted
observables from the  pure threshold values which is important
for the future understanding  of the role of an internal strangeness
degrees  of freedom in the nucleon. For this aim
we study the beam-target spin asymmetry and the relative role of the
singlet and triplet states in the entrance channel.

Our paper is organized as follows.
In Section II, we define the kinematical variables and formulae for
calculating the cross sections and polarization observables.
The basic amplitudes for the mechanisms illustrated in Figs.~1 and 2 are
given explicitly in Section III.
In Section IV we discuss  results of numerical
calculations and predictions.
The summary is given in Section V.
In the Appendix we describe the formalism for the enhancement
factors of final state interaction within the framework
of the Jost function and the effective phase-equivalent potentials.


\section{Observables}

The differential cross section of the reaction
$\pi^-p\to \phi n $ (cf. Fig.~1)
has the obvious form 
\begin{eqnarray}
\frac{d\sigma}{d\,\Omega_\phi}
=\frac{1}
{64\pi^2s}\,\frac{|{\bf q}|}{|{\bf p_\pi}|}\overline{|T_{(1)}|^2},
\label{CS_pi}
\end{eqnarray}
where $p_\pi=(E_\pi,\, {\bf p}_\pi) $ and $q=(E_\phi, {\bf q})$
are the four-momenta of the pion and the
$\phi$ meson in the center of mass system (c.m.s.);
$\overline{|T_{(1)}|^2}$ means average and sum over the initial and final
spin states, respectively.

The differential cross section of
$\phi$ production in the reaction $a+b\to c+d+\phi$, 
where $a,b$ and $c,d$ label the incoming and outgoing nucleons (cf. Fig.~2),
is related to the invariant amplitude $T_{(2)}$ as
\begin{eqnarray}
d\sigma =
\frac{1}{2(2\pi)^5\sqrt{s(s-4M_N^2)}} \,\overline{|T_{(2)}|^2} \,
\frac{d{\bf p}_c}{2E_c} \,
\frac{d{\bf p}_d}{2E_d} \,
\frac{d{\bf q}}{2E_\phi} \,
\delta^{(4)}(P_i - P_f).
\label{sig1}
\end{eqnarray}
where
$p_n=(E _n,\, {\bf p}_n)$ with $n=a,b,c,d$
are the four-momenta of the nucleons  in the c.m.s.,
$\sqrt{s}=E_a+E_b$
is the total c.m.s.\ energy,
$P_{i,f}$ are  the total four-momenta of the initial or final states.
Hereafter $\theta$ denotes the polar $\phi$ meson angle
and $\Omega$ is its solid angle.
We use a coordinate system with
${\bf z} \parallel {\bf p}_a$,
${\bf y} \parallel {\bf p}_a{\bf \times q}$.
Among the five independent variables for
describing the final state we choose
$E_\phi,\,\Omega\,$ and $\Omega_c$. Then the energy $E_c$
of particle $c$ is defined by
$E_c=\frac{AB - C\sqrt{B^2-M_N^2(A^2-C^2)}}{A^2-C^2},$
with
$A=2(\sqrt{s}-E_\phi)$,
$B=s-2E_\phi\sqrt{s}+M_\phi^2$,
$C=2|{\bf q}|\cos \theta_{qp_c}$,
Finally, the fivefold differential cross section reads
\begin{eqnarray}
\frac{d^5\sigma}{dE_\phi d\Omega d\Omega_c}
=
\frac{1}{8(2\pi)^5\sqrt{s(s-4M_N^2)}} \,
\overline{|T_{(2)}|^2} \,
\frac{|{\bf q}|\,|{\bf p}_c|^2}{|A\, |{\bf p}_c|+C E_c|}.
\label{dcs}
\end{eqnarray}
The total and/or partially differential cross sections are found
by integration over the available phase space.

In this paper we consider two polarization observables.
One of them is the spin density matrix which describes  the spin structure
of the outgoing $\phi$ meson,
\begin{eqnarray}
\rho_{rr'}
=
\frac{\sum_\beta T_{r,\beta }\, T^*_{r',\beta}}
{\sum_{r, \beta} T_{r,\beta} \,T^*_{r,\beta}},
\label{rhom}
\end{eqnarray}
where $r \equiv m_\phi = \pm 1, 0$ are the spin projections of the $\phi$
meson, and $\beta$ denotes a set of unobserved quantum numbers.
The spin density  defines the angular distribution in
$\phi\to e^+e^-$ and $\phi\to K^+K^-$ decays, which has a simple
form in a system where the $\phi$ meson is at
rest~(for details see~\cite{TKS}).
The decay angles $\Theta$, $\Phi$ are defined as polar and azimuthal
angles of the direction of flight of one of the decay particles
in the $\phi$ meson's rest frame.
The decay distributions integrated over the azimuthal angle $\Phi$,
${\cal W}(\cos\Theta)$,
depend only on the diagonal matrix elements $\rho_{00},\,
\rho_{11}=\rho_{-1-1}$, normalized as $\rho_{00}+ 2\rho_{11}=1$,
according to
\begin{eqnarray}
{\cal W}(\cos\Theta) & = &
\frac{3}{2(B +3 )}
\left(1+B\cos^2\Theta\right),
\label{W}
\end{eqnarray}
where the $\phi$ decay anisotropies $B$ read
\begin{eqnarray}
B^{K^+K^-} =
- \frac{1-3 \rho_{00}}{1-\rho_{00}}, \quad
B^{e^+e^-} = \frac{1 - 3 \rho_{00}}{1 + \rho_{00}}.
\label{B2}
\end{eqnarray}
To exclude the kinematical dependence of $\rho_{00}$ or $B$ on the
$\phi$ meson production angle, we choose the quantization axis along
the ${\bf z}$ direction (in the $\phi$ rest system),
and using the corresponding Wigner rotation functions
$d_{\cdots}^{\cdots}(\chi)$ one gets the amplitudes $T_{r,\beta}$ in
Eq.~(\ref{rhom}) by
\begin{eqnarray}
T^z_{m_{\phi},\, \beta}=\sum_{\lambda,\beta_i'}
T^{c.m.s.}_{\lambda,\, \beta'}\, d^1_{\lambda,m_\phi}(\chi_\phi)
\prod_{i}d^{\frac12}_{\beta_i',\beta_i}(\chi_i),
\label{Wigner}
\end{eqnarray}
where only $\chi_\phi=-\theta$ is important, while the
other $\chi_i$'s disappear in Eq.~(\ref{rhom}).

Another polarization observable is the beam-target asymmetry
in the $NN\to NN\phi$
reactions which is related to the nucleon spin states via
\begin{equation}
C_{BT} = \frac
{d\sigma(S_i=1) - d\sigma(S_i=0)}
{d\sigma(S_i=1) + d\sigma(S_i=0)},
\label{ASYM}
\end{equation}
where $S_i$ is the total spin in the entrance channel.
It is important to note that
spin and parity conservation arguments result in a
precise model independent prediction \cite{Rekalo} for the beam - target
asymmetry:
$ C_{BT} = 1$
for the $pp\to pp\phi$ reaction at the threshold.
In the  $pn\to pn\phi$ reaction the asymmetry depends on the relative
weights of the triplet and singlet states in the entrance channel.

\section{Basic amplitudes}

Basically, our consideration in this section is similar to the
previous study \cite{KO} (for the pure mesonic exchange contributions
depicted in Figs.~1a and 2a)
and to the models~\cite{TKS,Nakayama99} for both
channels shown in Figs.~1 and 2. The difference between
this work and previous ones is in the different form of cut-off
form factors for the off-shell nucleons in direct $\phi$ emission
(cf. Figs.~1b and 2b) and a different choice of the cut-off parameters
in $\pi N$ and $NN$ interactions which we will discuss below
in detail. In spite of the mentioned similarity, for completeness
in discussing our predictions for the set of observables
which have not been considered before, in this section
we display the main formulae which define the basic amplitudes.
The meson - nucleon and the $\phi\rho\pi$ interaction Lagrangians read
in  standard notation
\begin{eqnarray}
{\cal L}_{MNN} & = &
- i g_{\pi NN} \bar N \gamma_5 \bbox{\tau} \bbox{\pi} N\nonumber\\
&& - g_{\rho NN} \left( \bar N \gamma_\mu \bbox{\tau} N \bbox{\rho}^\mu
-\frac{\kappa_\rho}{2M_N} \bar N \sigma^{\mu\nu} \bbox{\tau}N
\partial_\nu\bbox{\rho}_\mu \right)\nonumber\\
&& - g_{\phi NN} \left( \bar N \gamma_\mu  N {\phi}^\mu
-\frac{\kappa_\phi}{2M_N} \bar N \sigma^{\mu\nu} N
\partial_\nu {\rho}_\mu \right),
\label{L_MNN} \\
{\cal L}_{\phi\rho\pi} & = &
g_{\phi\rho\pi} \, \epsilon^{\mu\nu\alpha\beta} \,
\partial_\mu \phi_\nu \, {\rm Tr} ( \partial_\alpha \rho_\beta \pi),
\label{L_PRP}
\end{eqnarray}
where  ${\rm Tr}(\rho \pi) = \rho^0 \pi^0 + \rho^+ \pi^- + \rho^- \pi^+$,
and bold face letters denote isovectors.
All coupling constants with off-shell meson are dressed by monopole
form factors \cite{Bonn}
$F_i=(\Lambda_i^2-m_i^2)/(\Lambda_i^2-k_i^2)$,
where $k_i$ is the four-momentum of the exchanged meson.
Following the scheme of the meson photoproduction \cite{Hab97} we assume
that $\phi NN$ vertices
must be dressed by form factors for off-shell virtual nucleons.
But this might result in  a  violation of the transversality
of the amplitude with respect to the $\phi$
meson field. To avoid this problem we use the prescription of
Ref.~\cite{Hab97} and parameterize the product of the two form factors
appearing in the left and the right diagrams in Figs.~1b and
2b in a symmetrical form
\begin{eqnarray}
F_{N}(p_l,p_r)=\frac12\left(
\frac{\Lambda_N^4}{\Lambda_N^4 + (p_l^2-M_N)^2 }
 + \frac{\Lambda_N^4}{\Lambda_N^4 + (p_r^2-M_N)^2 }
\right);
\label{cutN}
\end{eqnarray}
here  $p_l\,(p_r)$ is the four-momentum of the virtual nucleon in the
left (right)  diagrams in Figs.~1b and 2b.

\subsection{$\bbox{\pi N\to N\phi}$ reaction} 

The invariant amplitude for the meson exchange channel (a) in Fig.~1 reads
\begin{eqnarray}
T_{(1a) \, \lambda} = K^{\pi N} \,
\epsilon^{ijkl}
\left[ \bar u(p_c)\, \Gamma_{\rho \, l} (k_\rho) u(p_a) \right] \,
q_i k_k \, \varepsilon_j^{* \, \lambda} \, I_\pi,
\label{T_pi_M}
\end{eqnarray}
where
\begin{eqnarray}
\Gamma_\rho^i (k) = \gamma^i
 + i\frac{\kappa_\rho}{2M_N} \sigma^{ij}\,k_{\rho \, j},
\label{Gamma}
\end{eqnarray}
\begin{eqnarray}
K^{\pi N} (k_\rho) =
- \frac{g_{\rho NN}\,g_{\phi \rho\pi} }{k_\rho^2-m_\rho^2}
\frac{\Lambda_{\rho NN}^2-m_\rho^2}{\Lambda_{\rho NN}^2-k_\rho^2}
\frac{{\Lambda^{\rho \, 2}_{\phi\rho\pi}}-m_\rho^2}
{{\Lambda^{\rho 2}_{\phi\rho\pi}}-k_\rho^2}
\end{eqnarray}
with $k_\rho=p_c-p_a$ as the virtual $\rho$ meson's  four-momentum;
$\varepsilon_j^\lambda$ is the $\phi$ meson polarization vector,
$I_\pi$ denotes the isospin factor being equal 
to $\sqrt{2}$ (1) for a $\pi^-$
($\pi^0$) meson in the entrance channel,
and the nucleon spin indices are not displayed;
$i,j \cdots$ are Lorentz indices, and $\gamma_i$ and $u$ denote
Dirac matrices and bispinors.

The invariant amplitude for the direct radiation channel (b) in Fig.~1 has
the following form
\begin{eqnarray}
T_{(1b) \, \lambda} = i g_{\phi NN}\,g_{\pi NN}\,
\bar{u}(p_c)
\left[
\Gamma_\phi^i (-q) \,
\frac{\not\hskip-0.7mm\!{p}_l + M_N }{p_l^2-M_N^2}
+
\frac{\not\hskip-0.7mm\!{p}_r + M_N }{p_r^2-M_N^2}
\,\Gamma_\phi^i (-q)
\right]
\bar{u}(p_c) \, \varepsilon^{* \, \lambda}_i \, I_\pi \, F_N(p_l,p_r),
\label{T_pi_D}
\end{eqnarray}
where
$\Gamma_\phi^i (q)$ and
$F_N$ are defined by Eqs.~(\ref{Gamma}) and (\ref{cutN}), respectively,
and $p_l = p_a - q$ and $p_r = p_c - q$.

\subsection{$\bbox{NN\to NN\phi}$ reaction} 

The total invariant amplitude of meson exchange diagrams (a) in Fig.~2
with internal meson conversion
is the sum of 4 amplitudes
\begin{equation}
(T_M)_\alpha =
\xi_\alpha^1 T_M[ab;cd] + \xi_\alpha^2 T_M[ab;dc]
+\xi_\alpha^3  T_M[ba;dc]  + \xi_\alpha^4 T_M[ba;cd]
\label{T_NN_M}
\end{equation}
with
$\xi^1_{pp}=\xi^3_{pp}=-\xi^2_{pp}=-\xi^4_{pp}=1$,
$\xi^1_{pn}=\xi^3_{pn}=-1$,
$\xi^2_{pn}=\xi^4_{pn}=-2$.
The last two terms stem from the antisymmetrization or
from charged meson exchange in $pp$ or
$pn$ reactions, respectively\footnote
{In \cite{TKS} we used a convention with
$\xi^2_{pn}=\xi^4_{pn}=2$, which, however, does not change our
threshold prediction
for the ratio of the total cross sections in $pn$ and $pp$ interaction
made there without the final state interactions.}.
The first term in Eq.~(\ref{T_NN_M}) for the $pp$ reaction reads
\begin{eqnarray}
T_{(2a) \, \lambda}[ab;cd]=
K^{NN}\,
\left[
\bar u(p_d)\,\gamma_5\, u(p_b)\,\,
\right]\,
\left[
\bar u(p_c)\,\Gamma_\rho^j (k) u(p_a)
\, \epsilon_{ijkl}
k_{\rho}^i \,
q_\phi^k
\varepsilon^{* \, l}_\lambda
\right],
\end{eqnarray}
with
\begin{eqnarray}
K^{NN}(k_\pi^2,k_\rho^2) = -
\frac{g_{\pi NN} \, g_{\rho NN} \, g_{\phi\rho\pi}}
{(k^2_{\pi}-m_\pi^2)(k^2_{\rho}-m_\rho^2)}\,
\frac{\Lambda_\pi^2-m_\pi^2}{\Lambda_\pi^2-k^2_{\pi}}
\frac{\Lambda_\rho^2-m_\rho^2}{\Lambda_\rho^2-k^2_{\rho}}
\frac{{\Lambda^{\rho \, 2}_{\phi\rho\pi}}-m_\rho^2}
{{\Lambda^{\rho 2}_{\phi\rho\pi}}-k^2_{\rho}}
\frac{{\Lambda^{\pi \, 2}_{\phi\rho\pi}}-m_\pi^2}
{{\Lambda^{\pi 2}_{\phi\rho\pi}}-k^2_{\pi}}.
\label{K_factor}
\end{eqnarray}
The amplitude of direct $\phi$ meson
emission from the nucleon legs according to Fig.~2b
is calculated similarly
to the real or virtual photon
bremsstrahlung \cite{Giessen,TKB} in the few GeV region.
The internal zig-zag lines in Fig.~2b correspond
to a suitably parameterized $NN$ interaction
in terms of an effective two-body $T$-matrix which is written in the form
of the one-boson exchange  model (OBE) with effective coupling constants
and cut-off parameters and may be interpreted as effective
$\pi,\omega,\rho,\sigma$ meson exchanges.
We would like to stress
that this is an effective dynamical model
which is appropriate in the few GeV region and
which is different from the OBE model in the conventional sense.
This model has been applied successfully
to different reactions \cite{Giessen,TKB,Giessen2} and
this encourages us to employ it for the $\phi$ production too.

The total amplitude for the process (b) in Fig.~2 consists of $2 \cdot 8$
$(2\cdot 6)$ contributions for $pp$ ($pn$) interactions
and has a similar structure as
Eq.~(\ref{T_NN_M})
(with $\xi^1_{pn}=\xi^3_{pn}=1,\, \xi^2_{pn}=\xi^4_{pn}=0$,
for $\sigma,\omega$ exchanges), where $T[ab;cd]$ now reads
\begin{eqnarray}
&&T_{(2b) \, \lambda} [ab;cd]
= - g_{\phi NN}\,\varepsilon^{* \,\lambda}_i \,
\left[ \bar u(p_d)V^m u(p_b) \right] \times \nonumber\\
&&\sum_{m=\pi,\sigma,\rho,\omega}\,[-i\,D^m]
\,\bar u(p_c)\left[V^m\,
\frac{\not\hskip-0.7mm\!{p}_l + M_N }{p_l^2-M_N^2}
\,\Gamma_\phi^i (-q)
+ \Gamma_\phi^i (-q)\,
\frac{\not\hskip-0.7mm\!{p}_r + M_N }{p_r^2-M_N^2}
\,V^m
\right]u(p_a).
\label{T_NN_D}
\end{eqnarray}
Here, 
$V^m$ and $D^m$ are effective
coupling vertices and propagators of the two-body $T$ matrix, respectively,
\begin{equation}
D^{\pi, \sigma}=\frac{i}{k^2-m_{\pi, \sigma}^2}, \quad
{D_{\mu\nu}^{\rho, \omega}}=
-i\frac{g_{\mu\nu}- k_\mu k_\nu  m^{-2}_{\rho, \omega}}
{k^2-m_{\rho, \omega}^2},
\nonumber
\end{equation}
\begin{equation}
{V^{\pi}}= -iG_{\pi NN}\gamma_5,\,\,\,\,
{V^{\sigma}}= G_{\sigma NN},\,\,\,\,
V^i_{\rho, \omega}(k)=
- G_{\rho, \omega\,NN} \,\Gamma_{\rho, \omega}^i,
\label{POBE}
\end{equation}
where $k$ is the four momentum of the virtual meson $m$ and $G_{mNN}$ is the
vertex function which includes  the corresponding cut-off
form factor. The numerical values of the $G_{mNN}$ are taken from
Refs.~\cite{Giessen,Giessen2}.

In the near-threshold region the relative velocity of the outgoing nucleons
is small which might result in a strong final state
interaction (FSI) between them.
If the energy excess is a few MeV up to a few tens MeV
then one can consider only the
$s$-wave interaction and account for the final state interaction
in terms of the enhancement factors by renormalizing  the basic
amplitude correspondingly. For instance, for the $pn$ reaction we get
\begin{eqnarray}
{T}_{pn}[ab;cd]\to
{T}_{pn}[ab;cd]
\left({{\cal I}_0}_{pn}\,\delta_{-m_cm_d} +
{{\cal I}_1}_{pn}\,\delta_{m_cm_d}
\right),
\label{FSI_pn}
\end{eqnarray}
where $m_c,m_d$ are the spin projections of the nucleons in the final state,
and ${\cal I}_0,\,{\cal I}_1$ are the singlet and triplet
enhancement FSI factors, which are calculated within
the Jost function and the phase-equivalent potentials formalism,
which we describe in detail in Appendix A. The calculation shows
that the singlet enhancement factor is much greater than the triplet one, i.e.
$|{\cal I}_0|^2-1 \gg |{\cal I}_1|^2-1\simeq 0$ and greater than
the corresponding factors of higher partial waves. Thus, for the $pp$
interaction we use
\begin{eqnarray}
{T}_{pp}[ab;cd]\to
{T}_{pp}[ab;cd]
\left({{\cal I}_0}_{pp}\,\delta_{-m_cm_d} + \delta_{m_cm_d}
\right),
\label{FSI_pp}
\end{eqnarray}
reminding that at the threshold the $pp$ triplet final state
is exactly zero. $ {{\cal I}_0}_{pn}$ and
${{\cal I}_0}_{pp}$
are  different which reflects  the difference in the corresponding
effective radii and the scattering lengths.
Note that the mutual FSI of the $\phi$ and the outgoing nucleons is
assumed to be negligible.

In calculating the cross sections and the spin density matrix,
squares and bilinear forms of the FSI-corrected amplitudes need to be
evaluated numerically.


\section{Results}

\subsection{Fixing parameters}

The parameters of the two-body $T$ matrix for the direct $\phi$
emission depicted in Fig.~2b are taken from
Refs.~\cite{Giessen,Giessen2}, where a quite reasonable agreement with data
of different elastic and inelastic $NN$ reactions is found.

The coupling constant $g_{\phi\rho\pi}$
is determined by the $\phi\to\rho\pi$ decay.
The recent value $\Gamma( \phi\to\rho\pi) =$ 0.69 MeV
results in $|g_{\phi\rho\pi}|$ = 1.10 GeV$^{-1}$.
The SU(3) symmetry consideration \cite{Nakayama99,TLTS}
predicts a negative value for it.
Thus,  $g_{\phi\rho\pi}= - 1.10$ GeV$^{-1}$.

The remaining parameters of the meson exchange amplitudes for the processes
in Figs.~1a and 1b  are taken from the
Bonn model as listed in Table B.1 (Model II) of Ref.~\cite{Bonn}.

The yet undetermined  parameters are: the cut-off parameters for the virtual
mesons in the $\phi\rho\pi$ vertex, $\Lambda^\pi_{\phi\rho}$ and
$\Lambda^\rho_{\phi\rho}$, the cut-off $\Lambda_N$ in Eq.~(\ref{cutN}),
and the parameters of the $\phi NN$ interaction,
$g_{\phi NN}$ and $\kappa_\phi$.
We can reduce the  number of parameters by  making the natural assumption
$\Lambda^\pi_{\phi\rho}=\Lambda^\rho_{\phi\rho}$ based on the
symmetry of the virtual mesons in the $\phi\rho\pi$ vertex~\cite{Nakayama99}.
The  next consideration is related to the tensor coupling $\kappa_\phi$.
Based on the $\phi-\omega$ similarity we do not expect a large value for it
and in all our subsequent  calculations we employ the theoretical
estimate \cite{MMSV} $\kappa_\phi=0.2$ as an upper limit.

Even after that we have three free parameters being
$g_{\phi NN}$, $\Lambda^\rho_{\phi\rho}$ and $\Lambda_N$.
For $g_{\phi NN}$ the SU(3) symmetry predicts \cite{DeSwart}
\begin{eqnarray}
g_{\phi NN}=-{\rm tg}\,\Delta\theta_V\,g_{\omega NN},
\label{phi-omegaNN}
\end{eqnarray}
where  $\Delta\theta_V$ is the deviation from the ideal $\omega-\phi$
mixing angle. It is responsible for the ``standard'' OZI rule violation,
and in general, it depends on the method of its determination
(Gell-Mann--Okubo linear or quadratic mass formulae, radiative decays, say
$\phi(\omega)\to \gamma\pi$, etc.).  Using the quadratic
Gell-Mann--Okubo mass formula one gets  $\Delta\theta_V=3.7^0$.
Sometimes another relation is used, e.g.
\begin{eqnarray}
g_{\phi NN}=- 3 \sin\,\Delta\theta_V\,g_{\rho NN},
\label{phi-rhoNN}
\end{eqnarray}
which is obtained from the SU(3) relation \cite{DeSwart}
\begin{eqnarray}
g_{\omega NN}=\frac{3F-D}{F+D}\cos\Delta\theta_V\,g_{\rho NN},
\label{omega-rhoNN}
\end{eqnarray}
and Eq.~(\ref{phi-omegaNN}) with the assumption
$D/F=0$ in the SU(3) vector meson octet.
Using the known values for $g^2_{\rho NN}/4\pi=0.7-1.3$ \cite{Bonn}
and $g^2_{\omega NN}/4\pi=22-24$
\cite{Bonn}, one may obtain $-g_{\phi NN}=0.57-0.65$
and  $-g_{\phi NN}=1.07-1.09$ for the expressions (\ref{phi-rhoNN})
and (\ref{phi-omegaNN}),
respectively. On the other hand, the theoretical
estimates of Ref.~\cite{MMSV} give  $g_{\phi NN}=-0.24$.
Thus, we can conclude that even using the standard OZI rule violation
(thought non-ideal $\omega-\phi$ mixing)
one is left with estimated values of
$g_{\phi NN}$ within a quite large interval.
The possible hidden strangeness in a nucleon may even increase this interval.
In this paper we restrict ourselves to the standard OZI rule violation
mechanisms and
analyze consequences of varying $-g_{\phi NN}$ in the region 0.0 -- 1.0.

The negative coupling constant $g_{\phi\rho\pi}$
results in a destructive
interference between meson exchange amplitudes (a) and direct emission
(b)
in Figs.~1 and 2.
Analyzing the unpolarized $\pi^-p\to n\phi$ reaction, based on the data of
Ref.~\cite{DATApi}, we find  that the yet unconstrained three parameters
$g_{\phi NN}$, $\Lambda^\rho_{\phi\rho}$, $\Lambda_N$
become related to each other
as  $\Lambda^\rho_{\phi\rho}$=$\Lambda^\rho_{\phi\rho}
(g_{\phi NN},\,\Lambda^\rho_{\phi\rho})$
by the constrains given by the data,
and two solutions emerge
for this dependence: (i) $\sigma_{(a)} > \sigma_{(b)} $
and (ii)  $\sigma_{(a)} < \sigma_{(b)} $, where $\sigma_{(a,b)}$
are the total cross section for the meson exchange process (a) and the direct
$\phi$ emission (b) calculated separately.
These solutions are displayed
in Fig.~\ref{fig:3} for several values of $g_{\phi NN}$ as discussed above.

In order to constrain one more free parameter we analyze also the
cross section $d\sigma/d\Omega$ for the $pp\to pp\phi$ reaction, using
simultaneously the
DISTO data \cite{DISTO,DISTO99}, i.e.\
the angular distribution \cite{DISTO} (we remind that in our notation
$\Omega$ is the $\phi$ meson solid angle)
and the total cross section \cite{DISTO99}, respectively. For this aim
we fix the absolute
normalization of the angular distribution $d\sigma/d\Omega$
given in Ref.~\cite{DISTO} by making use of  the recently published
the total cross section \cite{DISTO99}.
As a result we get the fat dots
in Fig.~\ref{fig:3}. The minimum values for $-g_{\phi NN}$
are 0.07 (for $\sigma_{(a)}  > \sigma_{(b)}$) and
0.60 (for $\sigma_{(a)}  < \sigma_{(b)}$), respectively
at $\Lambda_N\to \infty$.
For both solutions we find that increasing values of $|g_{\phi NN}|$ results
in decreasing values of $\Lambda_N$ leaving the total cross section or
``integrated'' strength of the $\phi NN$ interaction on the same level.

To explore in more detail the relative importance of the
direct $\phi NN$ interaction we now employ three parameter sets;
two of them correspond to the $\sigma_{(a)}  > \sigma_{(b)}$
solution and the third one to $\sigma_{(a)}  < \sigma_{(b)}$:\\
set A: $g_{\phi NN}=-0.24$ \cite{MMSV},
$\Lambda^\rho_{\phi\rho}=1.34$ GeV, $\Lambda_N=1.065$ GeV
($\sigma_{(a)} >\sigma_{(b)} $),\\
set B: $g_{\phi NN}=-0.8$,
$\Lambda^\rho_{\phi\rho}=1.34$ GeV, $\Lambda_N=0.715$ GeV
($\sigma_{(a)} >\sigma_{(b)} $),\\
set C: $g_{\phi NN}=-0.8$,
$\Lambda^\rho_{\phi\rho}=1.60$ GeV, $\Lambda_N=1.99$ GeV
($\sigma_{(a)} <\sigma_{(b)} $).\\
In the sets B and C we choose $-g_{\phi NN}$ close to its upper limit
as predicted by SU(3) symmetry.

Results of our calculation  of the total cross section for
the $\pi^-p\to n\phi$ reaction for these parameter sets are shown in
Fig.~\ref{fig:4}
as a function of the energy excess
$\Delta s^{\frac12}\equiv s^{\frac12}-{s_0}_{\pi N}^{\frac12}$
with ${s_0}_{\pi N}=(M_N+m_\phi)^2$.
The contribution of the meson exchange channel (a)
is displayed separately by the
dot-dashed line, and the direct emission by the nucleon line (b)
is depicted by the dashed line.
Clearly seen is the strong destructive interference of the channels
(a) and (b), in particular for the set C.

In Figs.~\ref{fig:5} and~\ref{fig:6} we display results of our calculations
for $d\sigma/d\Omega$ for the $pp\to pp\phi$ reaction together with
the available data \cite{DISTO,DISTO99} 
for the parameter sets B and C, respectively,
at $\Delta s^{\frac12}= s^{\frac12}-{s_0}_{N N}^{\frac12} = 82$ MeV
with ${s_0}_{N N}=(2M_N+m_\phi)^2$.
The results for the set A are very similar to that of the set B
and we do not separately display them here.
Interestingly, in all parameter sets considered the channel (a) dominates
in the $NN \to NN \phi$ reaction, but the interferences are different for
different parameter sets.

One can see a qualitative difference in $\pi N$ and $NN$ reactions for
the set C. In the $\pi N$ reaction the direct radiation channel (b) is
dominant, i.e. $\sigma_{(b)} > \sigma_{(a)}$. In contrast, in the
$NN$ reaction the relative contribution of the direct emission channel
(b) increases considerably as compared with sets A and B, but it is
still smaller that the meson exchange channel (a). The reason for this
difference is the following one. The direct emission (b) in the
reaction $\pi N$ consists of the two competing u-channel and s-channel
diagrams shown in Fig.~1b. which add destructively, while the 
contribution of the of the u-channel amplitude is greater. However,
the corresponding contributions of the two competing diagrams in Fig.~2b  
are numerically nearly the same resulting in a stronger suppression
of the direct channel (b).

As we have adjusted our parameters by the data, it is clear that they
describe the data with  approximately
equal quality, and at  the present level of the data accuracy
it is difficult to give a preference to one of them.
Therefore,
we now investigate whether other observables can be
used to constrain the parameters further and whether the difference between
$pp$ and $pn$ reactions is a sensible measure.

\subsection{$\bbox{\pi^-p\to n\phi}$ reaction}

The calculated angular differential cross sections of the
$\pi^-p\to n\phi$ reaction at $\Delta s^{\frac12}=50$ MeV and for
the parameter sets A, B, C are shown in
Fig.~\ref{fig:7}.
One can see that  the shapes of the distributions
for the sets A and B are very similar to each other.
They are
quite smooth and close to the distribution of the meson exchange
channel (a). Only in the backward direction the total cross section slightly
decreases due to the destructive interference with the direct
channel (b), leading to some  enhancement of the cross section
in forward direction.
(We would like to mention here, that the extrapolated value of the
$\phi$ production cross section in $NN$ reactions \cite{Hibou} from the data
\cite{DISTO}  would require somewhat different parameter sets,
which in turn cause also more pronounced differences between the
sets A and B.)
Contrary
to that, for the model C the largest destructive interference appears
at forward direction, where the  contributions of the two competing channels
(a) and (b)
are close to each other. As a result,
the cross section is enhanced
in the backward direction. So we can conclude
that the differential cross section is sensitive to the dynamics
of the $\phi$ production and the direct $\phi NN$ coupling
(a similar conclusion for $\omega$ production has arrived at
in  Ref.~\cite{Nakayama98}).

The prediction for spin density matrix elements $\rho_{00}$ and
$\rho_{11}$ for the different models is shown in Fig.~\ref{fig:8}
as a function of the $\phi$
production angle in c.m.s.\ at
$\Delta s^{\frac12}=50$ MeV. The sets A and B deliver
standard values, typical for the spin-flip processes, i.e.
$\rho_{00}\simeq 0$, $\rho_{11}\simeq 0.5$. But the parameter set C predicts
a strong deviation from these values, especially at forward direction.
The reason of this effect is  the following. In the meson exchange
channel (a) the nucleon  spin-flip amplitudes result in transitions
$m_i\to m_\phi, m_f$, where $m_f=m_i-m_\phi$
with $m_\phi=\pm1$. For instance, the transitions
like $-\frac12\to -1, \frac{1}{2}$ are dominant. In  the direct
radiation channel (b) together with this strong amplitudes we have
finite amplitudes for the transition
$m_i\to 0, m_f$, where $m_f=m_i$. In the set C the strongly competing
nucleon spin flip amplitudes cancel
each other and only the nucleon spin conserving direct emission
amplitude (b) survives.
This is illustrated in Fig.~\ref{fig:9}, where we show
the nucleon spin flip (left panel) and the nucleon spin conserving
(right panel) amplitudes for the set C.
Here $F_{\rm z}={\rm Im} T^{\pi p\to n\phi}$ with the quantization axis
along the ${\bf z}$ direction  ($T^{\pi p\to n\phi}$ is purely imaginary).

The anisotropies of the decay channels
$\phi\to e^+e^-$ and  $\phi\to K^+K^-$
(cf.\ Eqs.~(\ref{W}, \ref{B2}) for
the different parameter sets are shown in Fig.~\ref{fig:10}. Again, one can
see a strong deviation of our prediction for the set C from the
naive expectation $B^{e^+e^-}\simeq 1$, $B^{K^+K^-}\simeq -1$ based on
a purely mesonic exchange channel or on the sets A and B.
Fig.~\ref{fig:11} illustrates the the manifestation of this deviation
in the real $e^+e^-$ and $K^+K^-$ angular distributions.
The distributions $W_L$ and $W_T$ are the
longitudinal (along the quantization axis) and transversal
fluxes for the outgoing electrons  or
kaons. The functions $W_L$ and $W_T$ are normalized as
$\int \sqrt{W_L^2(\Theta) + W_T^2(\Theta)}d\cos\Theta =1$, where
$\Theta$ is defined by Eq.~(\ref{W}).
Thus, one can see that the sets A and B predict a practically
vanishing   kaon flux in the longitudinal direction for all $\phi$
production angles.
The set C predicts a finite amount of the longitudinal
flux which increases with decreasing  $\phi$ production angle in
c.m.s.  A corresponding  modification is predicted for the electron flux too.

\subsection{$\bbox{pp\to pp\phi}$ and $\bbox{pn\to pn\phi}$ reactions} 

As shown in Figs.~\ref{fig:5} and \ref{fig:6}
the meson exchange contribution (a) is
the dominating contribution to the $NN\to NN\phi$ reaction,
therefore, it is useful to
recall the threshold prediction for this channel in the absence of the
final state interaction, which serves as a starting point for
further calculations at finite energy.
Adopting the notation of Ref.~\cite{Rekalo} we can express
the invariant amplitudes of the reaction $ab\to ab\phi$ with $a=p$, 
$b=p$ or $n$ as following
\begin{eqnarray}
T_{pp}=F_{1},\qquad\qquad
T_{pn}=\frac12\left(F_{0} +   F_{1}\right),
\label{THR1}
\end{eqnarray}
where $F_{0}\,(F_{1})$ is the initial singlet (triplet) amplitude with
\begin{eqnarray}
&&F_0=f_0(-1)^{\frac12+m_a}\,\delta_{-m_a m_b}
(\delta_{\frac{1}{2} m_c}\,\delta_{\frac{1}{2} m_d}
- \delta_{-\frac{1}{2} m_c}\,\delta_{-\frac{1}{2} m_d}),\nonumber\\
&&F_1=f_1(-1)^{\frac12+m_a}\,\delta_{m_am_b}
(\delta_{\frac{1}{2} m_c}\,\delta_{-\frac{1}{2} m_d}
- \delta_{-\frac{1}{2} m_c}\,\delta_{\frac{1}{2} m_d}),
\label{THR2}
\end{eqnarray}
where  $m_{a,b}$ and $m_{c,d}$ are again the nucleon spin projections
in the initial and final states, respectively, $f_0=6\sqrt{2} T_0$,
$f_1=2\sqrt{2} T_0$, where the threshold amplitude $T_0$ is defined by
Eq.~(25) in Ref.~\cite{TKS}. The above equations lead to the ratio
$f_0/f_1=3$ and to the ratio of singlet to triplet cross sections
\begin{eqnarray}
\frac{|F_0|^2}{|F_1|^2} = 9.
\label{ratioSTthr}
\end{eqnarray}
The beam target asymmetries~(\ref{ASYM}) read
${{C_{BT}}}_{pp}=1$, and ${{C_{BT}}}_{pn}=-0.8$.
The ratio of the total cross
sections in $pn$ and $pp$ reactions is 5. Accordingly, the prediction for
the spin density~\cite{Rekalo}
reads $\rho_{00}=0$,  $\rho_{11}=0.5$.

Let us now turn back to the Figs.~\ref{fig:5} and \ref{fig:6}. These
figures  show
(i) a relatively small contribution of the direct radiation channel (b),
which is in agreement with previous works~\cite{TKS,Nakayama99},
(ii) the cross sections for the $pn$ interaction are qualitatively very
similar in shape  to
these of the $pp$ interaction but they are larger, and (iii) the ratio of
the corresponding cross sections in $pn$ and $pp$ reactions
is different for the sets A (or B) and C. Later we will discuss
this aspect in  detail.

The energy dependence of the total cross sections of
$pp\to pp\phi$ and  $pn\to pn\phi$ reactions
for the sets B and C is shown in Figs.~\ref{fig:12} and \ref{fig:13}.
We do not display the result for the set A because it
is practically the same as for the set B.
The experimental data is taken from Ref.~\cite{DISTO99}.
One can see
that the direct radiation channel (b) is much indeed
smaller than the meson exchange
contribution (a) in the near-threshold region where our
consideration is valid.

Fig.~\ref{fig:14} shows the energy dependence of
the ratio of the total
cross sections of $pp\to pp\phi$ and $pn\to pn\phi$ reactions for
the different parameter sets.
One can see that this ratio  increases with the
energy excess and differs from the threshold value 5 in case of
absence of FSI.
The FSI  is greater in the triplet initial (or singlet
final) states and reduces the contribution of the initial singlet state
in the $pn$ interaction. For the set C the ratio
$\sigma_{pn}/\sigma_{pp}$
is greater because of the relatively greater
contribution of the initial
triplet state in the meson exchange channel (a). Fig.~\ref{fig:15}
shows the energy dependence of the ratio of the singlet to triplet
cross sections in $pn$ interactions (cf.\ Eq.~(\ref{ratioSTthr})).
The left panel shows this ratio  for the separate channels, while on the
right panel one can see our prediction for this ratio
for  the different parameters sets. One can again see
a strong deviation from the threshold prediction (\ref{ratioSTthr})
without FSI
and non-trivial non-monotonic dependence of these ratios
with some maximum values around $\Delta s^{\frac12}\sim 20$ MeV.

Fig.~\ref{fig:16} shows beam target asymmetry (\ref{ASYM}) for
the separate channels for $pp$ and $pn$
interactions. For the $pp$ interaction it coincides with its threshold value
$C_{BT}=1$ up to a relatively large energy excess.
For the $pn$ interaction
the asymmetry is different for the different channels which
reflects the different role of the singlet
and triplet states  in the different
amplitudes which are additionally modified by the FSI.

The total asymmetry for the different parameter sets is shown in
Fig.~\ref{fig:17}.  It is interesting that even for the sets A and B with
small contribution of the direct radiation amplitude (b) the asymmetry
for $pn$ interaction  strongly deviates
from the threshold prediction (without FSI: $C_{BT}^{pn}=-0.8$),
displaying a minimum  around $\Delta s^{\frac12}\sim 20$ MeV.

We do not display here our results for the spin density matrix
elements because for the sets A, B and C we get almost
the threshold values, i.e. $\rho_{00}=0$,  $\rho_{11}=0.5$, which reflects
the  dominance of the meson exchange channel (a).


\section{Summary}

We have analyzed the $\phi$ production
in $\pi N$ and $NN$ interactions in the near-threshold region
using the conventional ``non-strange'' hadron dynamics, based on the
amplitudes allowed by  non-ideal $\omega-\phi$ mixing, that is
meson conversion in a $\phi \pi \rho$ vertex
and direct $\phi$ emission from the nucleon legs
by a direct $\phi NN$ coupling.
Using the limited body of available experimental data
of the total unpolarized
reactions we have tried  to reduce as much as possible
the uncertainty of the model parameters. As a result we get two branches of
solutions with either a relatively small or a relatively large contribution
of the direct emission channel which is determined by  the
strength of the $\phi NN$ interaction. By making use of these
solutions we have compared various parameter sets with different strengths
of the  direct  $\phi NN$ interaction.

Analyzing the $\pi p\to n\phi $ reaction we find a strong dependence
of the various observables on the
strength of the $\phi NN$ interaction.
The study of the differential cross section and angular distributions
of electrons and kaons in the $\phi\to e^+e^-$ and $\phi\to K^+K^-$
decays seems  to be  most promising in investigating
the  $\phi NN$ dynamics. Experimentally, this study might be performed
with the pion beam at the HADES spectrometer in GSI/Darmstadt.

Analyzing the $NN\to NN\phi$ reaction we find
a large difference in $pp$ and $pn$  reactions due to the
different role of the singlet and triplet nucleon spin states
in the entrance channel and strong final state interaction.
We predict a non-monotonic energy dependence of the ratio
$\sigma_{pn}/\sigma_{pp}$ and of the beam target asymmetry for the  $pn$
interaction which deviates strongly from the pure threshold
prediction.

Finally, we emphasize once more that the present investigation is completely
based on the conventional meson-nucleon dynamics and,
therefore, our predictions may be considered as a necessary
background for forthcoming studies of the strangeness
degrees of freedom in  non-strange hadrons.
Additionally we would like to mention that fixing the $\phi N N$
coupling is important for an access to the elastic $\phi N$ scattering
cross section which determines the degree of thermalization and 
collective flow properties of the $\phi$ mesons in heavy-ion
collisions at SIS energies. 

\subsection*{Acknowledgments}

We gratefully acknowledge  fruitful discussions with
M. Debowski, L.P. Kaptari, N. Kaiser, R. Kotte, J. Ritman, and V.V. Shklyar.
One of the authors (A.I.T.) thanks for the warm hospitality
of the nuclear
theory group in the Research Center Rossendorf.
This work is supported by BMBF grant 06DR829/1,
Heisenberg-Landau program, and
HADES-JINR participation project \#03-1-1020-95/2002.


\appendix

\section{Final state interaction}

In this appendix we present  the formulae for the FSI
and corresponding correction factors. We use the general framework
for the  FSI enhancement factor based on the Jost function
formalism. Important aspects of this framework are
described in the monograph by Gillespie \cite{Gill64}
and some early  original papers \cite{Early}.
With respect to the
significance of this problem in studying various near-threshold
particle production reactions in the present time with cooled beams,
we accumulate
here the relevant expressions of this method
and give the final result in a  form convenient for specific
calculations. For the Jost function formalism we use the notation
of the textbook by Newton \cite{Newt_book}.
For simplicity, we limit
our consideration to the  $s$-wave interaction which is dominant
in the near-threshold region. A generalization for higher angular
momenta  may be done straightforwardly.

The FSI enhancement factor for two identical particles with momentum
$k$ in their c.m.s.\ reads
\begin{eqnarray}
{\cal I}=\,\frac{1}{{\cal J}_+(k)}\,\, ,
\label{A1}
\end{eqnarray}
where ${\cal J}_{+}$ belongs to a set of functions  ${\cal J}_{\pm}(k)$
which are  defined through the Wronskian of two linearly
independent
solutions of the Schr\"odinger equation,
\begin{eqnarray}
{\cal J}_{\pm}(k)=f_{\pm}(k,r)\,\varphi'(k,r)
- f_{\pm}'(k,r)\,\varphi(k,r),
\label{A2}
\end{eqnarray}
where the prime means here the derivative with respect to $r$.
The function ${\cal J}$= ${\cal J}_+$ is called the Jost function.
The integral equations for the regular and irregular functions
$\varphi(k,r) $ and $f_{\pm}(k,r)$
 have the standard form
\begin{eqnarray}
\varphi(k,r)=\frac{\sin kr}{k}  +\frac{1}{k}\int_0^\infty dr'\,
\sin k(r-r')V(r')f_{\pm}(k,r'),\nonumber\\
f_{\pm}(k,r)={\rm e}^{\pm kr} -\frac{1}{k}\int_0^\infty dr'\,
\sin k(r-r')V(r')f_{\pm}(k,r').
\label{A3}
\end{eqnarray}
Eq.~(\ref{A2}) and the boundary conditions
$\varphi(0)=0,\,\varphi'(0)= 1 $ show that
${\cal J}_{\pm}(k)=f_+(k,0)$, thus  allowing the integral
representation
\begin{eqnarray}
{\cal J}(k)=1 +\frac{1}{k}\int_0^\infty dr\, \sin kr V(r) f_+(k,r).
\label{A4}
\end{eqnarray}
The physical wave function $\psi^+(k,r)$ is related to the regular
function $\varphi(k,r)$ as
\begin{eqnarray}
\psi^+(k,r)=\frac{k\varphi(k,r)}{{\cal J}(k)},
\label{A5}
\end{eqnarray}
which means that the inverse of the square of
the modulus of ${\cal J}(k)$
measures the probability of finding the particles near $r=0$,
relative to a situation without interaction.
From Eq.~(\ref{A4})
one can find the important asymptotic condition
\begin{eqnarray}
\lim_{|k| \to \infty}{\cal J} =1,
\label{A6}
\end{eqnarray}
which shows  that at high energies the enhancement tends
to unity, thus leaving  the total amplitude  unchanged.

The analyticity  of ${\cal J}$ together with the asymptotic condition
Eq.~(\ref{A6}) leads to
the integral representation of the Jost function in terms
of the phase shift
$\delta(k)$
\begin{eqnarray}
{\cal J}(k)=\prod_{n}\left(1+\frac{\kappa_n^2}{k^2}\right)
\exp\left[\frac{1}{\pi}\int_{-\infty}^{\infty}
dk'\frac{\delta(k')}{k-k'+i\epsilon}  \right],
\label{A7}
\end{eqnarray}
where $\kappa_n$ is related to the corresponding binding
energy $\kappa_n^2=-2\mu\epsilon_n>0$ if bound states appear;
$\mu$ is the corresponding reduced mass.

For the practical usage of the above formalism it is convenient
to work with the effective potentials which give an exact analytical
expression for the phase shift.
Let us first consider
the singlet $NN$ scattering (without bound state). The Eckart
potential
\begin{eqnarray}
V(r) = -\frac{8\alpha^2}{\alpha^2-\beta^2}
\left(
\frac{{\rm e}^{-\alpha r}}{\alpha - \beta}
+ \frac{{\rm e}^{\alpha r}} {\alpha  + \beta}
\right)^{-2},\,\,\,\,\,\, \alpha >0, \beta >0,
\label{A8}
\end{eqnarray}
gives the $s$-wave phase shift
\begin{eqnarray}
k\, {\rm cot}\,\delta_0 =
\frac{\alpha\beta}{\alpha -\beta}
 + \frac{k^2}{\alpha -\beta}
\label{A9}
\end{eqnarray}
reproducing  the effective-range  phase shift exactly,
\begin{eqnarray}
k\, {\rm cot}\,\delta_0 =
-\frac{1}{a_0}
+ \frac{1}{2}r_0{k^2},
\label{A10}
\end{eqnarray}
with
\begin{eqnarray}
\alpha=\frac{1}{r_0}\left(\sqrt{1-2 r_0 a_0^{-1}} +1 \right),\,\, \,\,\,
\beta=\frac{1}{r_0}\left(\sqrt{1-2 r_0 a_0^{-1}}  -1 \right).
\label{A11}
\end{eqnarray}
The insertion  of $\delta_0$ from (\ref{A9}),
\begin{eqnarray}
\delta_0=\frac{i}{2}{\rm ln}\left[
\frac{(k-i\alpha)(k + i\beta)}{(k+i\alpha)(k -  i\beta)}
\right],
\label{A12}
\end{eqnarray}
in Eq.~(\ref{A7}) gives the analytical expression for the Jost function
\begin{eqnarray}
{\cal J}(k)=\frac{k + i\beta}{k+i\alpha},
\label{A13}
\end{eqnarray}
and the resulting  enhancement factor for singlet interaction
reads therefore as
\begin{eqnarray}
&&{\cal I}_0(k)= C_0(k^2)
\,\frac{\sin\, \delta_0\,{\rm e}^{i\delta_0}}{k},\nonumber\\
&&C_0(k^2)=\frac{k^2 +\alpha^2 }{\alpha-\beta}=
\frac{(kr_0)^2 + 2 \left( 1- r_0 a_0^{-1}
+ \sqrt{1- 2 r_0 a_0^{-1} } \right) }{2r_0},
\label{A14}
\end{eqnarray}
which coincides with the classical Watson enhancement
factor \cite{Watson52}
\begin{eqnarray}
{\cal I}_W(k)= C_W
\,\frac{\sin\, \delta_0\,{\rm e}^{i\delta_0}}{k}
\label{A15}
\end{eqnarray}
in the limit of $kr_0\to 0$. Note that expression (\ref{A15}) is
commonly used in calculations of the FSI in the near-threshold
region (cf.~\cite{Kaizer99}), 
where the constant $C_W$ is fixed
by a comparison of calculation and  experimental data.
Eq.~(\ref{A14}) is
superior to Eq.~(\ref{A15}) because it yields a  definite
value of $C_W= C_0(k^2=0)$ by making use of the independent
phase shift data and satisfies at the same time the
required asymptotic behavior according to Eq.~(\ref{A6}),
\begin{eqnarray}
\lim_{|k| \to \infty}{\cal I}_0(k) =1,
\label{A16}
\end{eqnarray}
contrary to Eq.~(\ref{A15}), where $ {\cal I}_W(k)\to 0$ at
${|k|\to \infty}$.

Fig.~{\ref{fig:FSI}} illustrates the effect of FSI
when using different enhancement factors for our parameter set B
for the  reaction $pp\to pp\phi$,
where we keep here only the dominant mesonic exchange diagram in Fig.~2a
and use the threshold value. In this case the energy
dependence comes only from the enhancement factor and
the phase space  volume.
The cross section calculated with enhancement factors from
Eqs.~(\ref{A14}) and (\ref{A15})
with $C_W=C_0(0)$ are  shown by the dashed and the solid lines,
respectively. For comparison we show also
the cross section calculated  without FSI (dot-dashed line).
One can see, that the difference between the two
factors (\ref{A14}) and (\ref{A15}) is indeed negligible
at sufficiently small energy excess, say at  $\Delta s^{\frac12}<10$ MeV with
$kr_0 \ll 1$, where the Watson theory \cite{Watson52}
is valid, thus supporting the approach of \cite{Kaizer99}.
At $\Delta s^{\frac12} > 50$ MeV
the difference between the two variants
results in a factor 2 and greater.

Following  Ref.~\cite{Newt_book},
for the triplet $pn$ interaction
one can use the effective potential of the Bargmann type
\begin{eqnarray}
&&V(r) = -4\kappa\frac{d}{dr}
\left[
{\rm sh}\,\alpha_1 r \frac{g(\kappa,r)}
{g(\kappa + \alpha_1,r) - g(\kappa - \alpha_1,r)}
\right],\nonumber\\
&&g(q,r)=\left({\rm e}^{-qr} + 2{\rm sh}\,qr
\right) k^{-1}
\label{A17}
\end{eqnarray}
which reproduces the known phase shift and the deuteron binding
energy. The Jost function in this case reads
\begin{eqnarray}
{\cal J}(k)=\frac{k - i\kappa}{k+i\alpha_1},
\label{A18}
\end{eqnarray}
with
\begin{eqnarray}
\alpha_1=(2-\kappa r_1) r_1^{-1},
\label{A19}
\end{eqnarray}
where $\kappa^2=2\mu \epsilon_d$, $\epsilon_d$ is the deuteron
binding energy, and  $r_1$ and $a_1=(\kappa(1-\kappa r/2))^{-1}$
are the triplet effective radius and  scattering length,
respectively. The triplet enhancement factor reads explicitly
\begin{eqnarray}
&&{\cal I}_1(k)= C_1(k^2)
\,\frac{\sin\, \delta_1\,{\rm e}^{i\delta_1}}{k},\nonumber\\
&&C_1(k^2)=\frac{k^2 +\alpha^2_1 }{\alpha_1 +\kappa}=
\frac{(kr_1)^2 + 2 \left(1- r_1 a_1^{-1}
+ \sqrt{1- 2 r_1 a_1^{-1} } \right) }{2r_1},
\label{A20}
\end{eqnarray}

In our calculation we use $a_{0,1}$ and $r_{0,1}$  from
Ref.~\cite{Dumbr}
\begin{eqnarray}
&& pn\,\,\, {\rm singlet}: \,\,\,\,\,a_{0pn}=-23.768\,{\rm fm},
\,\,\,\,\,
r_{0pn}=2.75\,{\rm fm},\nonumber\\
&& pp\,\,\, {\rm singlet}: \,\,\,\,\,a_{0pp}=-7.8098\,{\rm fm},
\,\,\,\,\,
r_{0pp}= 2.767\,{\rm fm},\nonumber\\
&& pn\,\,\, {\rm triplet}: \,\,\,\,\,a_{1pn}=\,\,\,\,\,
5.424\,{\rm fm}, \,\,\,\,\,\,\,\,r_{1pn}=1.759\,{\rm fm},
\,\,\,\, \kappa^{-1} = 4.318 \,{\rm fm}.
\label{A21}
\end{eqnarray}

Finally, we would like to mention that the approach presented here
is equivalent to the approach in Ref.~\cite{FSI} if
the off-shell correction factor ${\cal P}$ (cf.\ Eq.~(9) in \cite{FSI})
takes the form
${\cal P}(E,k) = - a_x r_x^{-1}
\left( 1 + \sqrt{1 - 2 r_x a_x^{-1}} \right)$
with $x = 0,1$.

\newpage

\begin{figure}
\centering{\epsfig{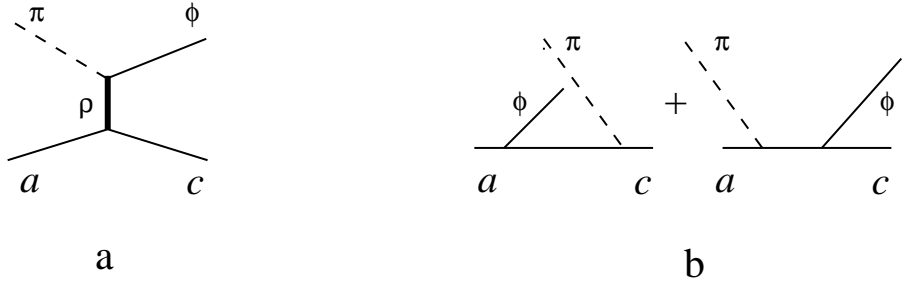}}

\vspace*{1.5cm}
\caption{
Diagrammatic representation of the $\pi N\to N\phi$ reaction mechanisms:
(a) meson exchange diagram with
$\phi$ emission from the $\phi\rho\pi$ vertex,
(b) direct $\phi$ emission from the $\phi NN$ vertex
in Compton like diagrams.
}
\label{fig:d1}
\end{figure}

\vspace*{1cm}

\begin{figure}
\centering{\epsfig{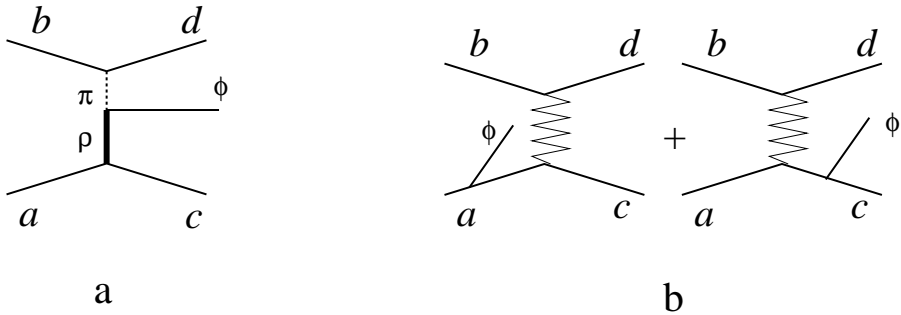}}

\vspace*{0.5cm}
\caption{
Diagrammatic representation of the $NN\to NN\phi$ reaction mechanisms:
(a) meson exchange diagram with
$\phi$ emission from the internal meson conversion in the
$\phi\rho\pi$ vertex,
(b) direct $\phi$ emission from nucleon legs. The zig-zag lines depict
effective boson exchange.
Exchange diagrams are not displayed.
}
\label{fig:d2}
\end{figure}

\newpage
\begin{figure}
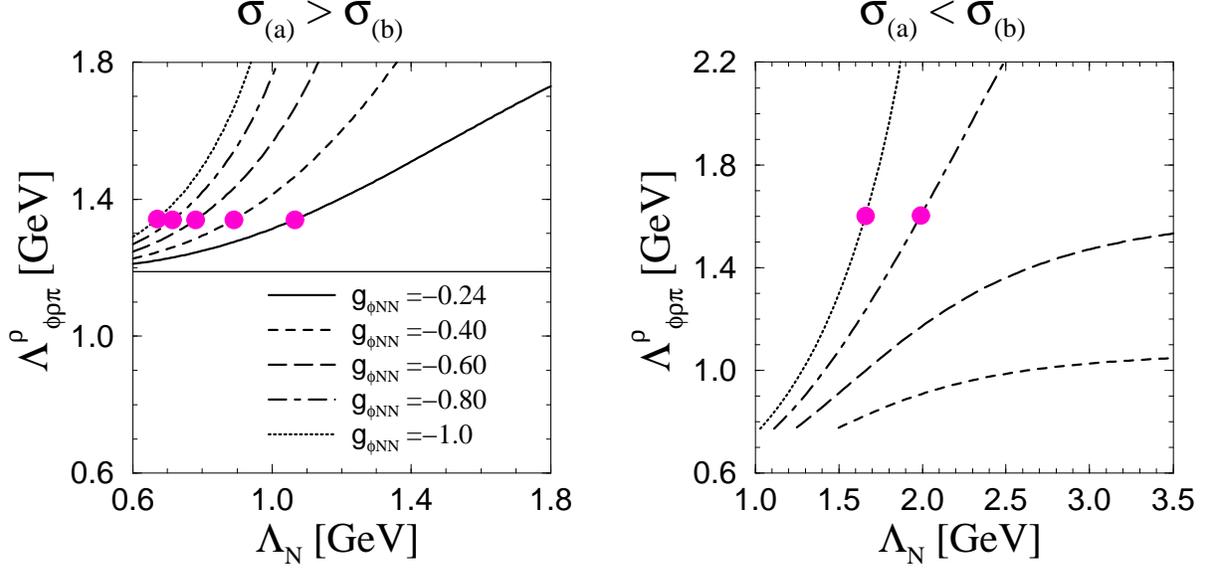

\centering{\epsfig{file=cutplus.epsi, width=7.5cm}
           \hspace*{5mm}
	   \epsfig{file=cutmins.epsi, width=7.5cm}}
\vspace*{0.5cm}
\caption{
Solutions for
the function $\Lambda^\rho_{\phi\rho}$=$\Lambda^\rho_{\phi\rho}
(g_{\phi NN},\,\Lambda^\rho_{\phi\rho})$;
left panel: $\sigma_{(a)} > \sigma_{(b)}$,
right panel: $\sigma_{(a)} < \sigma_{(b)}$.
The thin straight solid line in the left panel corresponds to $g_{\phi NN}=0$.
Further explanations are given in the text.
}
\label{fig:3}
\end{figure}

\begin{figure}
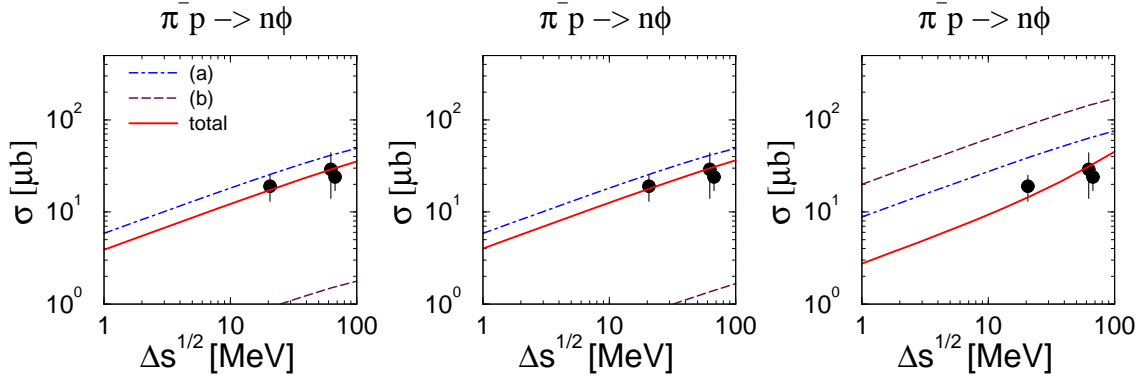

\centering{\epsfig{file=pi_tot024.epsi,  width=4.9cm}
           \epsfig{file=pi_tot080.epsi,  width=4.9cm}
	   \epsfig{file=pi_tot080_.epsi, width=4.9cm}}
\vspace*{0.5cm}
\caption{
The total cross section for the
$\pi^-p\to n\phi$ reaction for the parameters sets A, B, C
(left, middle, right panels)
as a function of the energy excess $\Delta s^{\frac12}$.
Data from \protect\cite{DATApi}.
}
\label{fig:4}
\end{figure}

\begin{figure}
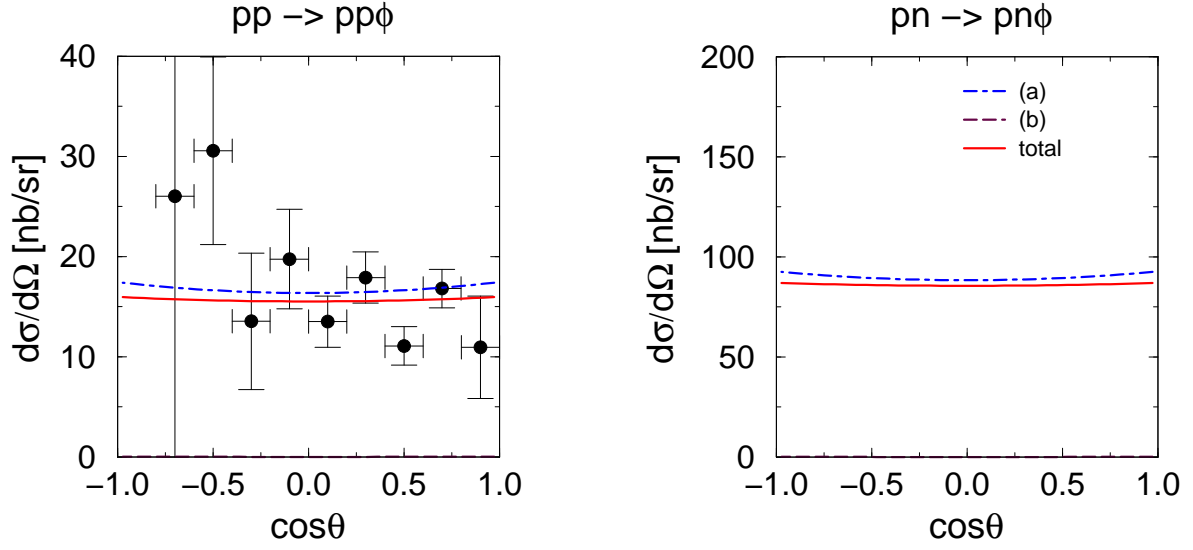

\centering{\epsfig{file=CS_TETH_L80.epsi,  width=6.77cm}\qquad\qquad
           \epsfig{file=CS_TETH_L80pn.epsi ,  width=7.1cm}}

\vspace*{0.5cm}
\caption{
The angular distribution
$d\sigma/d\Omega$ for $NN\to NN\phi$ reactions for parameter set B for
$pp$ (left panel; data  from \protect\cite{DISTO} with normalization
according to \protect\cite{DISTO99})
and $pn$ (right panel) interactions.
}
\label{fig:5}
\end{figure}


\begin{figure}
\centering{\epsfig{file=CS_TETH_L80_.epsi,  width=6.77cm}\qquad\qquad
           \epsfig{file=CS_TETH_L80_pn.epsi ,  width=7.1cm}}

\vspace*{0.5cm}
\caption{
The same as in Fig.~\ref{fig:5}, but for parameter set C.
}
\label{fig:6}
\end{figure}

%
%
%


\begin{figure}
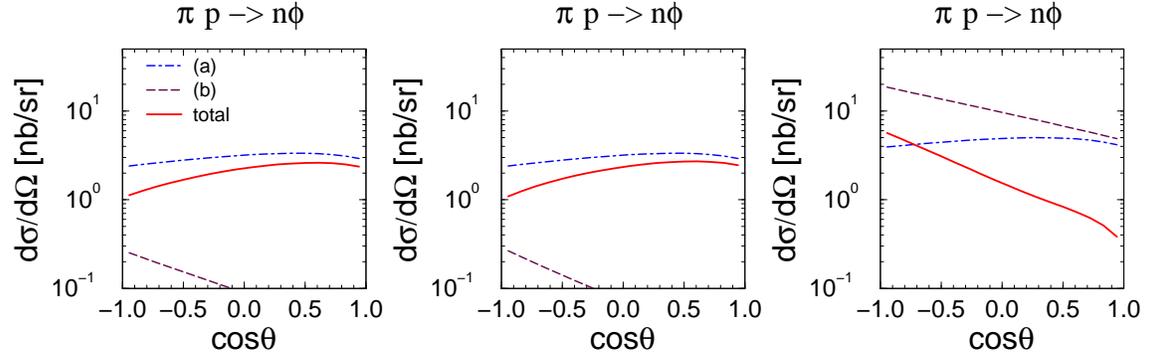

\centering{\epsfig{file=pi_t24.epsi,  width=4.9cm}
           \epsfig{file=pi_t80.epsi,  width=4.9cm}
	   \epsfig{file=pi_t80_.epsi, width=4.9cm}}

\vspace*{0.5cm}
\caption{
The differential cross sections for the
$\pi^-p\to n\phi$ reaction at $\Delta s^{\frac12}=50$ MeV
for the same parameter sets as in Fig.~\ref{fig:4}.
}
\label{fig:7}
\end{figure}


\begin{figure}
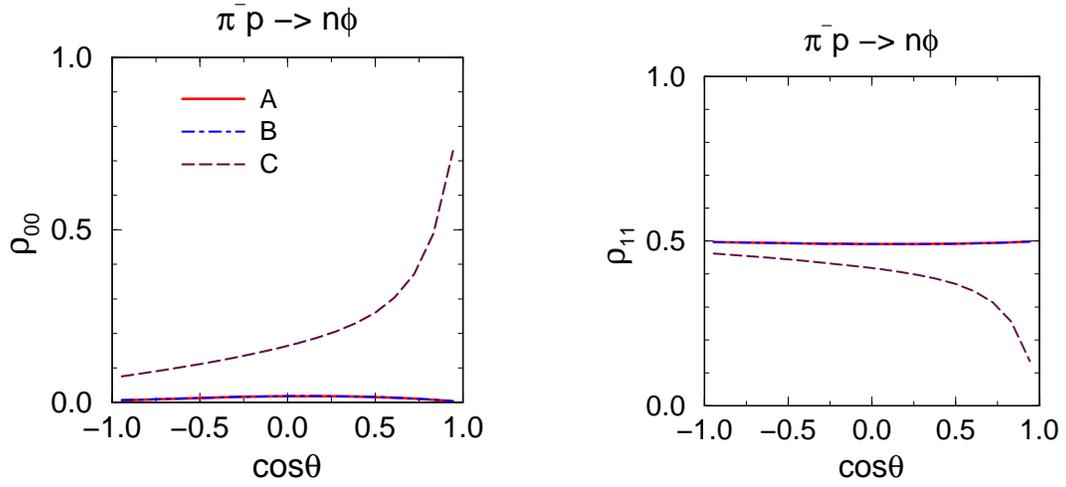

\centering{\epsfig{file=pi_rho_00.epsi,  width=6.286cm}\qquad\qquad
           \epsfig{file=pi_rho_11.epsi,  width=6cm}}

\vspace*{0.5cm}
\caption{
The spin density matrix elements $\rho_{00}$ (left panel) and
$\rho_{11}$ (right panel) for the different parameter sets at
$\Delta s^{\frac12}=50$ MeV.
}
\label{fig:8}
\end{figure}

\newpage

\begin{figure}
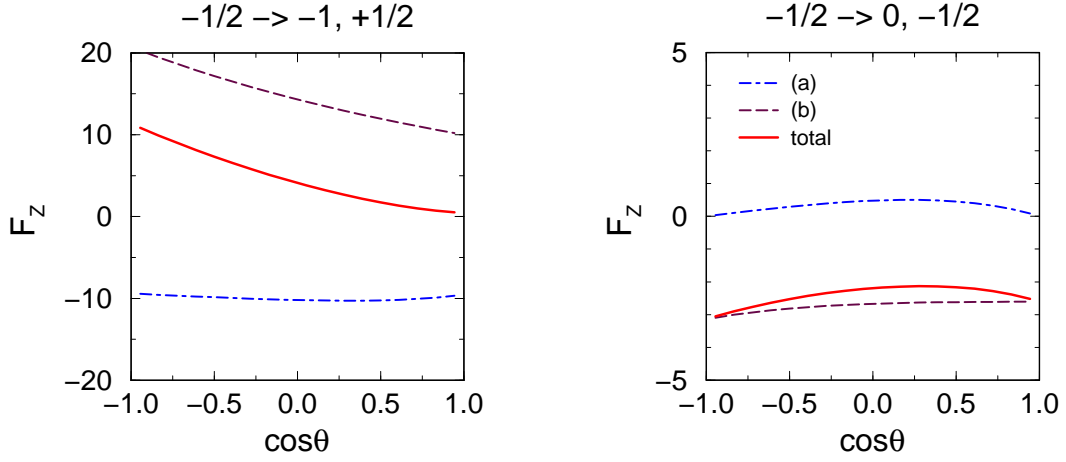

\centering{\epsfig{file=Am-Mp.epsi,  width=6.286cm}\qquad\qquad
           \epsfig{file=Am-Om.epsi,  width=6cm}}

\vspace*{0.5cm}
\caption{
The nucleon spin flip (left panel) and spin conserving (right panel)
amplitudes $F_z$ for the parameter set C.
}
\vspace*{0.5cm}
\label{fig:9}
\end{figure}

\begin{figure}
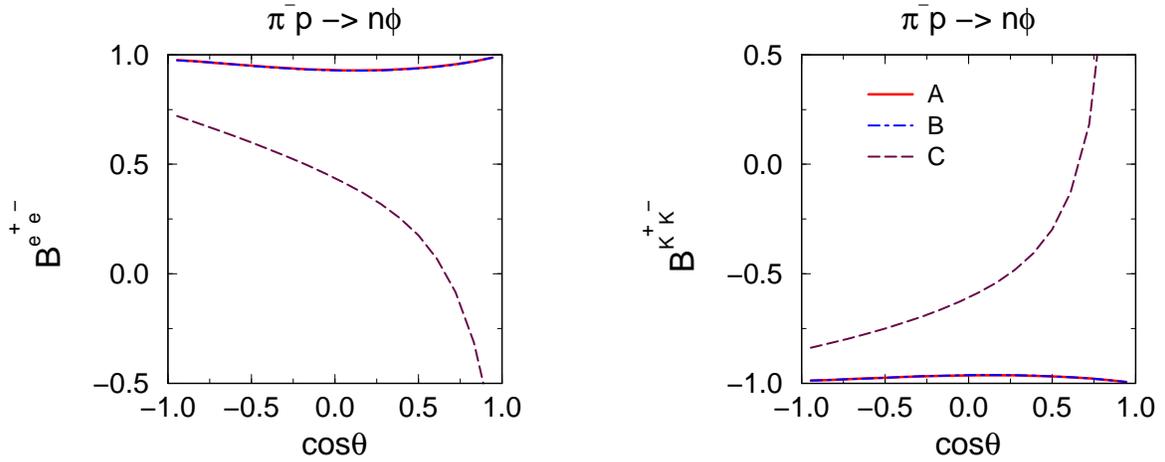

\centering{\epsfig{file=pi_bee.epsi,  width=6.77cm}\qquad\qquad
           \epsfig{file=pi_bkk.epsi,  width=6.77cm}}

\vspace*{0.5cm}
\caption{
The anisotropies in the reactions  $\phi\to e^+e^-$ (left panel)
and  $\phi\to K^+K^-$ (right panel)  for different parameter sets.
}
\label{fig:10}
\end{figure}

\begin{figure}
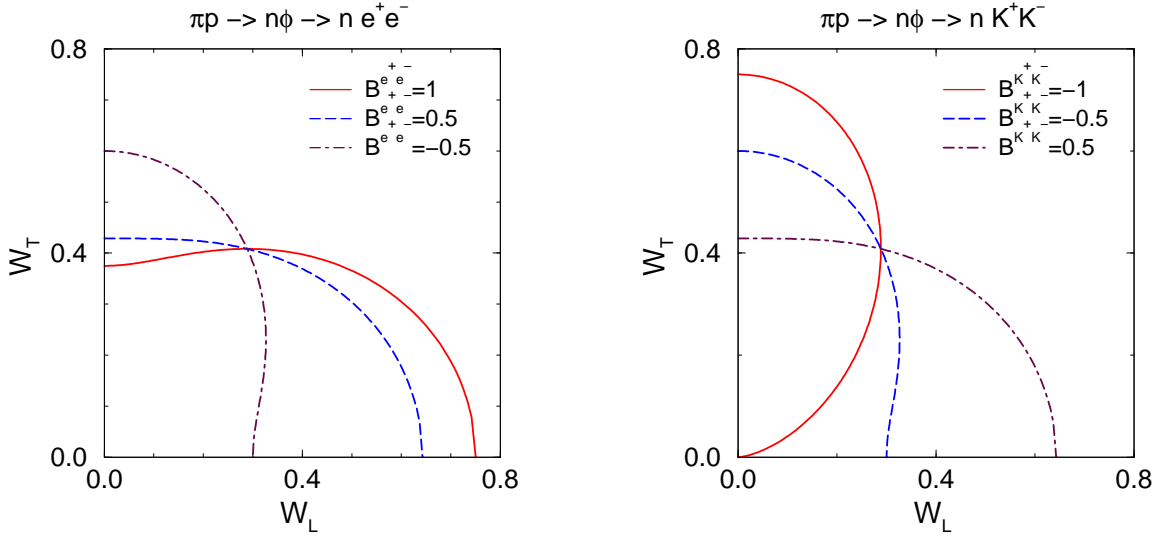

\centering{\epsfig{file=angle_ee.epsi,  width=6.77cm}\qquad\qquad
           \epsfig{file=angle_kk.epsi,  width=6.77cm}}

\vspace*{0.5cm}
\caption{
The longitudinal  ($W_L$)
and transversal ($W_T$) fluxes
of the outgoing electrons (left panel) or
kaons (right panel) for various values of the anisotropies.
}
\label{fig:11}
\end{figure}

%
%
%

\begin{figure}
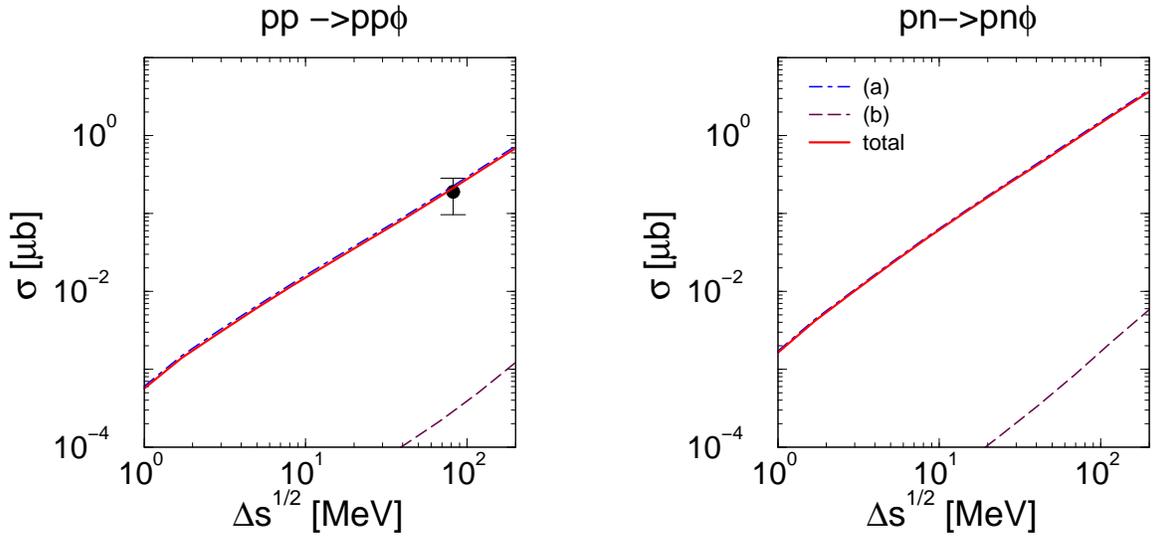

\centering{\epsfig{file=CS_S_pp80.epsi,   width=6.77cm}\qquad\qquad
           \epsfig{file=CS_S_pn80.epsi ,  width=6.77cm}}

\vspace*{0.5cm}
\caption{
The energy dependence of the total cross sections of
$pp\to pp\phi$ (left panel) and  $pn\to pn\phi$ (right panel)
reactions for the parameter set B.
Data from \protect\cite{DISTO99}.
}
\label{fig:12}
\end{figure}


\begin{figure}
\centering{\epsfig{file=CS_S_pp80_.epsi,   width=6.77cm}\qquad\qquad
           \epsfig{file=CS_S_pn80_.epsi ,  width=6.77cm}}

\vspace*{0.5cm}
\caption{
The same as in Fig.~\ref{fig:12}, but  for the parameter set C.
}
\label{fig:13}
\end{figure}

\begin{figure}
\centering{\epsfig{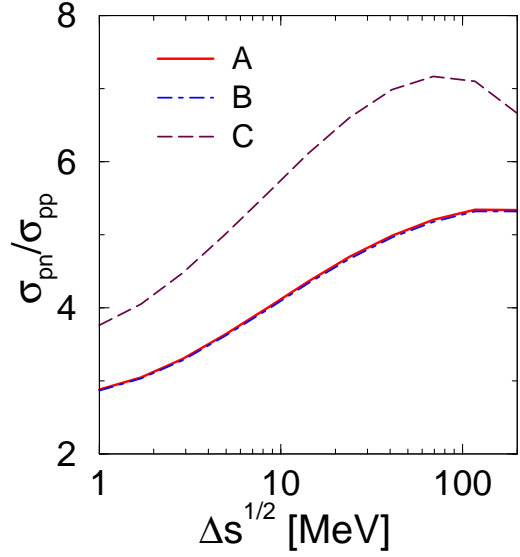}}

\vspace*{0.5cm}
\caption{
The energy dependence of the ratio of the total
cross sections of $pp\to pp\phi$ and $pn\to pn\phi$ reactions for
different parameter sets.
}
\label{fig:14}
\end{figure}


\begin{figure}
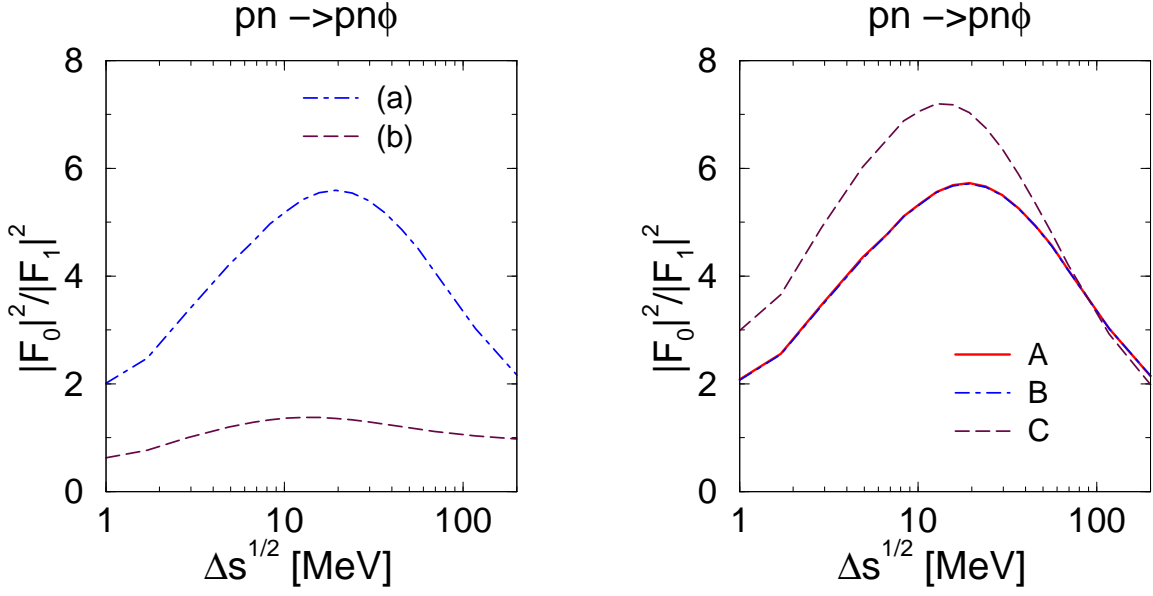

\centering{\epsfig{file=F_ST_80.epsi, width=6.77cm}\qquad\qquad
           \epsfig{file=F_ST.epsi,  width=6.77cm}}

\vspace*{0.5cm}
\caption{
The energy dependence of the ratio of the initial singlet to triplet
cross sections in $pn$ interaction
for separate channels (left panel) and for different
parameter sets (right panel).
}
\label{fig:15}
\end{figure}


\begin{figure}
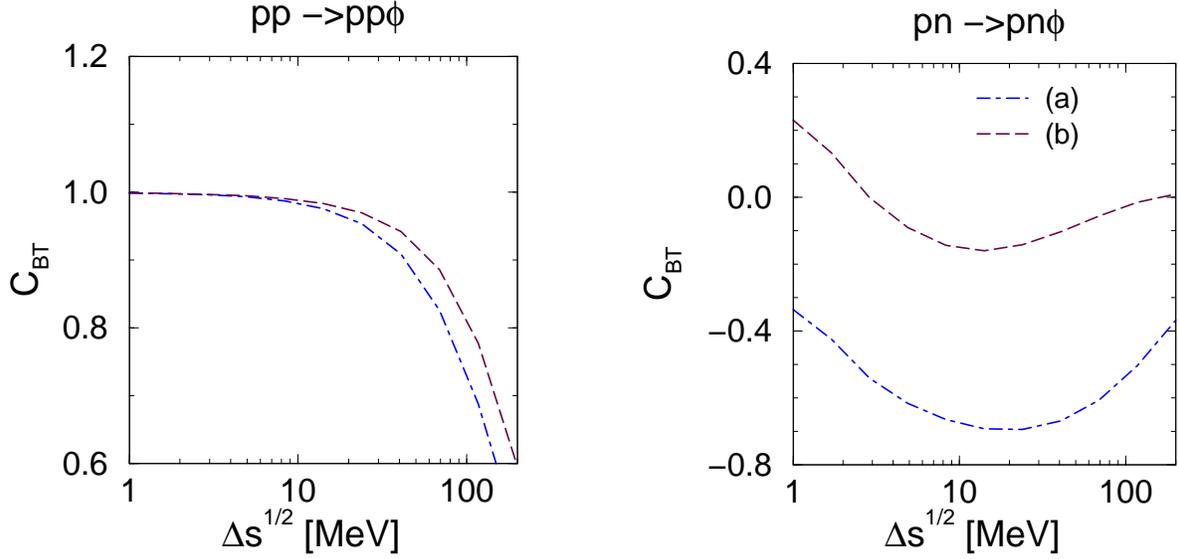

\centering{\epsfig{file=AS_pp_80.epsi,   width=6.77cm}\qquad\qquad
           \epsfig{file=AS_pn_80.epsi ,  width=7.1cm}}

\vspace*{0.5cm}
\caption{
The beam target asymmetry  for
separate channels for $pp$ (left panel) and $pn$ (right panel)
interactions.
}
\label{fig:16}
\end{figure}

\begin{figure}
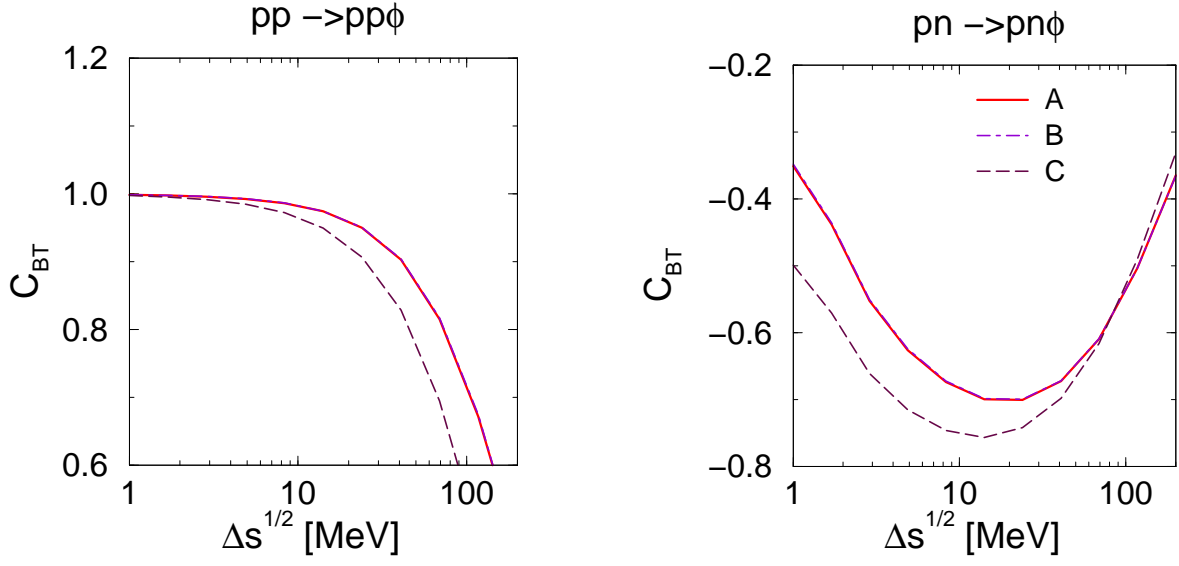

\centering{\epsfig{file=AS_pp.epsi,   width=6.77cm}\qquad\qquad
           \epsfig{file=AS_pn.epsi ,  width=7.1cm}}

\vspace*{0.5cm}
\caption{
The same as in Fig.~\ref{fig:15}
but for  different parameters sets.
}
\label{fig:17}
\end{figure}


\begin{figure}
\centering{\epsfig{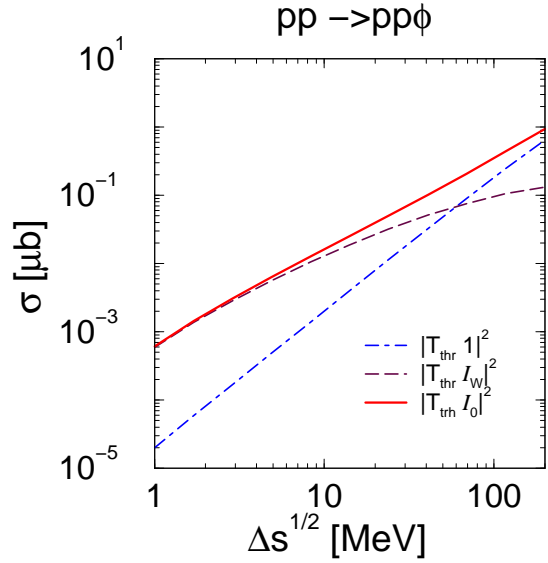}}

\vspace*{0.5cm}
\caption{
The effect of FSI for  different enhancement factors for the parameter set B
in the reaction $pp\to pp\phi$.
Notation is explained in the  text.
}
\label{fig:FSI}
\end{figure}


\begin{thebibliography}{99}

\bibitem{Ellis}
J. Ellis, M. Karliner, D.E. Kharzeev and  M.G. Sapozhnikov,
Phys. Lett. B {\bf 353}, 319 (1995).

\bibitem{OZI}
S. Okubo, Phys. Lett. B {\bf 5} (1963) 165;\\
G. Zweig, CERN report No. 8419/TH 412 (1964);\\
I. Iizuka, Prog. Theor. Phys. Suppl. {\bf 37/38} (1966) 21.

\bibitem{LZL93}
M.P. Locher and Y. Lu, Z. Phys. A {\bf 351}, 83 (1995);\\
D. Buzatu and  F.M. Lev,  Phys. Lett. B {\bf 329}, 143 (1994).

\bibitem{sigmaterm}
J. Gasser, H. Leutwyler and  M.E. Sainio, Phys. Lett. B {\bf 253}, 252 (1991).

\bibitem{newexp}
D.B. Kaplan and  A.V. Manohar, Nucl. Phys. B {\bf 310} 527 (1988);\\
R.D. McKeown, Phys. Lett. B {\bf 219}, 140 (1989);\\
E.M. Henley, G. Krein, S.J. Pollock and A.G. Williams,
Phys. Lett. B {\bf 269}, 31 (1991).

\bibitem{TOY97_98}
A.I. Titov, Y. Oh and S.N. Yang, Phys. Rev. Lett. {\bf 79}, 1634 (1997);\\
A.I. Titov, Y. Oh, S.N. Yang and T. Morii,
Phys. Rev. C {\bf 58}, 2429 (1998).

\bibitem{Ellis99}
J. Ellis, M. Karliner, D.E. Kharzeev and  M.G. Sapozhnikov,
hep-ph/9909235.

\bibitem{KO}
W.S. Chung, G.Q. Li and C.M. Ko, Phys. Lett. B {\bf 401}, 1 (1997).

\bibitem{TKS}
A.I. Titov, B. K\"ampfer, V.V. Shkyar,
Phys. Rev. C {\bf 59}, 999 (1999).

\bibitem{DISTO}
F. Balestra et al. (DISTO Collaboration),
Phys. Rev. Lett.  {\bf 81}, 4572 (1998).

\bibitem{DISTO99}
F. Balestra et al. (DISTO Collaboration),
Phys. Lett. B  {\bf 468}, 7 (1999).

\bibitem{Kotte}
N. Herrmann (FOPI collaboration), Nucl. Phys. A 610 (1996) 49c,\\
A. Mangiarotti, invited talk at the FOPI collaboration meeting,
Obernai, France, Sep. 29 - 30, 1999,\\
R. Kotte, private communication

\bibitem{HADES}
J. Friese et al. (HADES collaboration), GSI report 97-1, p. 193 (1997).

\bibitem{DATApi}
A. Baldini et al.,
{\it Total cross sections of high energy particles},
Springer-Verlag, Heidelberg, 1988.

\bibitem{Nakayama99}
K. Nakayama, J.W. Durso, J. Haidenbauer, C.Hanhart and J. Speth,
Phys. Rev. C {\bf 60}, 055209 (1999).

\bibitem{Hibou}
F. Hibou et al.,
Phys. Rev. Lett. {\bf 83}, 492 (1999).

\bibitem{Rekalo}
M.P. Rekalo, J. Arvieux and  E. Tomasi-Gustafsson,
Z. Phys. A {\bf 357}, 133 (1997).

\bibitem{Bonn}
R. Machleidt, Adv. Nucl. Phys. {\bf 19}, 189 (1989).

\bibitem{Hab97}
H. Haberzettl,
Phys. Rev. C  {\bf 56}, 2041 (1997);\\
H. Haberzettl, C. Bennhold, T. Mart and  T. Feuster,
Phys. Rev. C {\bf 58}, 40 (1998).

\bibitem{Giessen}
M. Sch\"afer, H.C. D\"onges, A. Engel and  U. Mosel,
Nucl. Phys. A {\bf 575}, 429 (1994).

\bibitem{TKB}
A.I. Titov, B. K\"ampfer and E. Bratkovskaya,
Phys. Rev. C {\bf 51}, 227 (1995).

\bibitem{Giessen2}
A. Engel, R. Shyam, U. Mosel and  A.K. Dutt-Mazumder,
Nucl. Phys. A {\bf 603}, 387 (1996);\\
V.V. Shklyar, B. K\"ampfer, B.L. Reznik and A.I. Titov,
Nucl. Phys. A {\bf 628}, 255  (1998).

\bibitem{TLTS}
A.I. Titov, T.-S.H. Lee, H. Toki and O. Streltsova,
Phys. Rev. C {\bf 60}, 035205 (1999).

\bibitem{MMSV}
U.-G. Meissner, V. Mull, J. Speth and J.W. Van Orden,
Phys. Lett. B {\bf 408}, 381 (1997).

\bibitem{DeSwart}
J.J.~De Swart,
Rev. Mod. Phys.  {\bf 35}, 916 (1963).


\bibitem{Nakayama98}
K. Nakayama, A. Szczurec, C. Hanhart, J. Haidenbauer and J. Speth,
Phys. Rev. C {\bf 57} 1580 (1998).


\bibitem{Gill64}
J. Gillespie,
  {\it Final-State Interactions},
  Holden-Day Advanced Physics Monographs, 1964.

\bibitem{Early}
R. Jost and W Kohn,
Phys. Rev. {\bf 87}, 977 (1952);\\
V. Bargmann,
Rev. Mod. Phys. {\bf 21}, 488 (1949);\\
E. Fermi,
Suppl. Nuovo Cimento,
{\bf 2}, 17 (1955);\\
R.G. Newton,
Phys. Rev. {\bf 105}, 763 (1957); {\bf 107}, 1025 (1957).

\bibitem{Newt_book}
R.G. Newton,
  {\it Scattering Theory of Particles and Waves},
  McGrow-Hill Inc., 1966.

\bibitem{Watson52}
K.M. Watson, Phys. Rev. {\bf 88}, 1163 (1952);
  {\it Scattering Theory of Particles and Waves},
  McGrow-Hill Inc., 1966.

\bibitem{Kaizer99}
N. Kaiser, Phys. Rev. C {\bf60}, 057001 (1999);\\
V. Bernard, N. Kaiser and  Ulf-G. Meissner,
Eur. Phys. J. A {\bf 4}, 259 (1999).

\bibitem{FSI}
C. Hanhart and K. Nakayama, Phys. Lett. B {\bf 454}, 176 (1999).

\bibitem{Dumbr}
O. Dumbrajs, R. Koch, H. Pilkuhn,
G.C. Oades, H. Behrens, J.J. De Swart and  P. Kroll,
Nucl. Phys. B {\bf 216}, 277 (1983).


\end{thebibliography}
\end{document}